 \documentclass[preprint2]{aastex}
\usepackage{rotating}
\usepackage{lscape}

\shorttitle{Infrared Models of Super Star Clusters}
\shortauthors{Whelan, Johnson, Whitney, Indebetouw, \& Wood}

\begin{document}

\title{The Infrared Properties of Embedded Super Star Clusters:
Predictions from Three-Dimensional Radiative Transfer Models}

\author{David G. Whelan}
\affil{Department of Astronomy, University of Virginia, P.O. Box 400325,
    Charlottesville, VA 22904}
\email{dww7v@astro.virginia.edu}

\author{Kelsey E. Johnson\altaffilmark{1}}
\affil{Department of Astronomy, University of Virginia, P.O. Box 400325,
    Charlottesville, VA 22904}
\email{kej7a@virginia.edu}

\author{Barbara A. Whitney}
\affil{Department of Astronomy, University of Wisconsin-Madison, 475 N
Charter Street, Madison, WI 53706}
\email{bwhitney@astro.wisc.edu}

\author{R\'{e}my Indebetouw\altaffilmark{1}}
\affil{Department of Astronomy, University of Virginia, P.O. Box 400325,
    Charlottesville, VA 22904}
\email{remy@virginia.edu}

\and

\author{Kenneth Wood}
\affil{School of Physics and Astronomy, University of St. Andrews, North Haugh,
St. Andrews, Fife KY16 9AD, Scotland, UK}
\email{kw25@st-andrews.ac.uk}


\altaffiltext{1}{Adjunct at National Radio Astronomy Observatory, 520
Edgemont Road, Charlottesville, VA 22904, USA}

\begin{abstract}

With high-resolution infrared data becoming available that can probe
the formation of high-mass stellar clusters for the first time,
appropriate models that make testable predictions of these objects are
necessary. We utilize a three-dimensional radiative transfer code,
including a hierarchically clumped dusty envelope, to study the
earliest stages of super star cluster evolution. We explore a range of
parameter space in geometric sequences that mimic the hypothesized
evolution of an embedded super star cluster. The inclusion of a
hierarchically clumped medium can make the envelope porous, in
accordance with previous models and supporting observational
evidence. The infrared luminosity inferred from observations can
differ by a factor of two from the true value in the clumpiest
envelopes depending on the viewing angle. The infrared spectral energy
distribution (SED) also varies with viewing angle for clumpy
envelopes, creating a range in possible observable infrared colors and
magnitudes, silicate feature depths and dust continua. General
observable features of cluster evolution differ between envelopes that
are relatively opaque or transparent to mid-infrared photons. For
optically thick envelopes, evolution is marked by a gradual decline of
the $9.8$\micron~ silicate absorption feature depth and a
corresponding increase in the visual/ultraviolet flux. For the
optically thin envelopes, clusters typically begin with a strong hot
dust component and silicates in emission, and these features gradually
fade until the mid-infrared PAH features are predominant. For the
models with a smooth dust distribution, the {\em Spitzer} MIPS or {\em
Herschel} PACS [$70$]-[$160$] color is a good probe of the stellar
mass relative to the total mass, or star formation
efficiency. Likewise, the IRAC/MIPS [$3.6$]-[$24$] color can be used
to constrain the R$_{in}$ and R$_{out}$ values of the
envelope. However, clumpiness confuses the general trends seen in the
smooth dust distribution models, making it harder to determine a
unique set of envelope properties. Nevertheless, good diagnostic
colors were found for each of the input parameters: again, the
[$70$]-[$160$] color can be used to separate models with different
star formation efficiencies; the {\em Spitzer} IRAC/MIPS
[$8.0$]-[$24$] color is able to constrain R$_{in}$ and R$_{out}$
values; and the IRAC [$3.6$]-[$5.8$] color is sensitive to the
fraction of the dust distributed in clumps. Finally, in a comparison
of this model set to IRAS data of ultracompact H{\sc ii} regions, we
find good agreement, suggesting that these models are physically
relevant, and will provide useful diagnostic ability for datasets of
resolved, embedded SSCs with the advent of high-resolution infrared
telescopes like {\em James Webb Space Telescope}.

\end{abstract}
 

\section{INTRODUCTION}

Super star clusters (SSCs) are massive young star clusters with high
stellar densities ($\gtrsim 10^3$ stars pc$^{-3}$) that form under
extreme pressures, often found in merging galaxy systems, galactic
nuclei, and blue compact dwarf galaxies \citep[][and references
therein]{whitmore02}. With masses typically in excess of $10^5
M_\sun$, they are the most massive type of stellar cluster
known. Studies of the most nearby analogues Westerlund $1$
\citep{clark05} and R$136$ \citep{bosch09} show that these objects are
consistent with expectations that they are globular cluster
progenitors; this expectation is strengthened by numerous studies of
SSC populations outside the Local Group \citep[e.g.][]{schweizer96,
holtzman92, oconnell94, conti94, johnson99}. Because of their high
density of massive stars, models show that SSCs have the ability to
disperse metals from supernovae ejecta to large distances, trigger
further star formation episodes, and act as a launching mechanism for
super-galactic winds \citep{wunsch08, murray10b}. Therefore SSCs can
have a major impact on their host galaxies.

Tracking the early evolution of SSCs requires long wavelength
observations due to the dust-enshrouding birth envelopes that surround
them at young ages. Embedded SSCs were first detected in the radio
\citep[e.g.][]{kobulnicky99,turner00} in low-metallicity blue compact
dwarf galaxies (BCDs) as compact sources with high ionizing
fluxes. Numerous other examples have also been found in the infrared,
in starbursts such as Arp $220$ \citep{shioya01} and again in BCDs
\citep{sauvage03}.

Radio observations have confirmed the masses of embedded SSCs to be in
the range found for their optical counterparts. Furthermore, the
calculated high ionizing fluxes are equivalent to hundreds or
thousands of O-stars and electron densities are $\sim
10^{3-4}$cm$^{-3}$ \citep{beck02,johnson03a,johnson03b}.  These
densities are similar to those found in Galactic HII regions, but are
significantly lower than typical densities associated with Active
Galactic Nucleus Broad Line Regions ($> 10^{8}$cm$^{-3}$) suggesting
that SSCs are also heated by young stars in a starburst environment,
and not AGN.

Despite advances made this last decade in understanding embedded SSCs 
as starburst events, their early evolution is still poorly
understood. Properties such as the mass and physical size of the
clouds from which they are formed and the amount of gas that is turned
into stars \citep[i.e. the star formation efficiency;][]{ashman01} are
not well-constrained. In addition, the most recent radio,
near-infrared, and ultraviolet data on these objects suggest that the
thick envelopes are porous, allowing a significant fraction of UV
light to leak from the system \citep{reines08, johnson09,
thuan05}. This project is a first attempt at modeling the infrared
spectral energy distributions of embedded super star clusters. By
investigating a large variation in the values of the input parameters
in a hierarchically clumpy and porous envelope, these models can be
used to constrain the dusty envelope geometry and star formation
efficiency of embedded super star clusters.

The decision to model embedded SSCs in the infrared is based on the
fact that they are most visible in the infrared and radio; at these
young ages, hot stars heat the dust in the circum-cluster envelope and
excite free-free radio emission inside the H{\sc ii} region. Therefore
these wavelengths are essential for studying the dust properties
during the embedded phase, the dust mass, and the envelope
geometry. The radiative transfer models presented in this work offer
predictions about what we expect embedded SSCs to look like in the
infrared and what wavelength observations offer the best diagnostic
capability. With telescopes like {\it JWST} on the horizon, its high
spatial resolution and mid-infrared sensitivity making it far better
than existing infrared space facilities \citep{sonneborn08}, we may
finally be entering a period when questions about initial dust mass
and geometry of a SSC's embedded phase can be answered on a global
scale. In the near term, observations using {\it Herschel} and {\it
Spitzer} may be used to probe unresolved embedded SSCs, and these
models can be used to estimate their physical size and geometry.

In this paper we present simulated infrared images, SEDs, and colors
of embedded SSCs along a geometric sequence that mimics the evolution
of a young embedded super star cluster. We discuss the models in
detail in \S\ref{models}, then present model images, SEDs and colors
in \S\ref{results}. We compare unresolved populations to the models in
\S\ref{comparison}, discuss the limitations of these models when
comparing to resolved populations in \S\ref{limitations}, and conclude
in \S\ref{conclusions}.

\section{MODELS \label{models}}

\subsection{The Radiative Transfer Method \label{rt}}

The models presented in this paper calculate radiative transfer from
thermal dust grains and stochastic small grains, and scattering from
the dust. The code simulates a three-dimensional geometry, and is
based on previous work on thermal dust grains in two dimensions
\citep[see][]{whitney03,chakrabarti09}, updated to include three
dimensions and clumpy dust distributions \citep{indebetouw06}, and
stochastic emission from polycyclic aromatic hydrocarbons and very
small grains \citep[PAHs/VSGs;][]{wood08}.

The code uses a Monte-Carlo technique that issues `photon packets'
into a dusty envelope from a central source. Packets that encounter
large grains either change the temperature in the grid space in which
the encounter occurs, according to the prescription of \citet{lucy99},
or are scattered according to a modified Henyey-Greenstein function
\citep{cornette92}. A small number of packets encounter PAHs/VSGs, and
are re-emitted according to emissivity templates from
\citet{draineli07}; see \citet{wood08} for details about emission from
PAHs/VSGs in our models.

Embedded SSCs are believed to have porous envelopes that allow a
significant portion of the ultraviolet (UV) light to escape from the
system \citep[supporting observational evidence is provided in
\S\ref{ObsEvidence} from][]{thuan05, reines08, johnson09}. To account
for this fact, we have included a hierarchically clumped density
structure for the dusty envelope, using the prescription of
\citet{elmegreen97}.

In order to confine the parameter space investigated in this model
set, we have varied only four input parameters that most influence the
output spectral energy distribution: the inner and outer dust radii in
a spherical geometry; the mass in stars divided by the total cluster
mass \citep[i.e. the `star formation efficiency' or SFE;][]{ashman01};
and the fraction of dust that is smoothly, as opposed to fractally,
distributed. The many other parameters that remain fixed, such as the
central star cluster mass, the IMF, the central source luminosity and
age, the dust species used, the volume fractal dimension adopted, and
the radial dust distribution exponent, are all described in
\S\ref{dust} to \S\ref{envelope}.

\subsection{The Dust Composition \label{dust}}

We use the standard dust model size distribution derived in
\citet{kim94} that fits an extinction curve with R$_V=4$ using the
prescription of \citet{cardelli89}. The size distribution is not a
simple power law, but is derived from a Maximum Entropy Method
solution, which gives the best solution compatible with the extinction
data. The chosen R$_V$ value has been found to be appropriate for the
more dense regions of molecular clouds \citep{whittet01} and therefore
we make the assumption that it will also fit embedded SSCs.

We adopt the dielectric functions of astronomical silicate and
graphite grains from \citet{laor93}. Our grain model includes a layer
of water ice on the grains, covering the outer $5$\% in radius. The
ice mantle increases the opacity of the grains at all wavelengths, but
the change is most pronounced longward of 35\micron. The slope of the
opacity with wavelength is the same in the infrared for grains with
and without the ice mantle, so any error introduced by the inclusion
of an ice mantle is a scalar factor at the wavelengths of concern in
this paper. The ice dielectric function is from \citet{warren84}.

The PAHs/VSGs are templates from \citet{draineli07}. The energy
density of the radiation field relative to the interstellar radiation
field (ISRF) can vary between $0.5 \leq U \leq 10^7$ and the
ionization fraction is that used in \citet{draineli07} to constrain
the models to reproduce the observed Milky Way spectrum. Only grains
that are $\leq 200$\AA~ in size are considered small grains. Grains
larger than this are assumed to thermalize after photon interactions,
and are therefore considered large grains. The PAH/VSG mass fraction
in dust used in this paper is $3.55$\%, which is the average value
found for $65$ galaxies in the SINGS survey \citep{draine07}. This is
therefore an acceptable value to expect for extragalactic embedded
SSCs.

\subsection{The Central Source \label{CentralSource}}

In these models, the central cluster is treated as a point source with
the spectrum of a $10^6$ M$_\sun$ star cluster, produced by the
Starburst99 spectral synthesis models \citep{leitherer99}. The input
metallicity is solar and the initial mass function is a Salpeter IMF
\citep{salpeter55}; changing these parameters was found to have little
effect on the overall shape of the resulting infrared SED, validating
the decision to hold them constant. The age is set to $1$ Myr
throughout the geometric sequence to ensure that the UV luminosity
remains constant. Since evolution is expected to take place in less
than three or four million years, prescribing a fixed age is warranted
- see \S\ref{ObsEvidence} for a discussion about the evolutionary pace
for embedded super star clusters. The instantaneous starburst mode was
chosen, without continuum emission. The luminosity of the source is
$1.6 \times 10^9$L$_\sun$.

\subsection{The Envelope Properties \label{envelope}}

The dust and gas around the central source is distributed spherically
with variable inner and outer radii and envelope mass. The initial
envelope mass is set for star formation efficiencies (SFEs) of $5$\%,
$10$\%, $15$\%, $25$\%, and $50$\%, which is a large range of possible
values based on observations of star-forming regions \citep{bonnell10,
murray10a}. The dust can be distributed fractally and smoothly, and
the percentage of dust that is smoothly distributed is varied from
between $100$\% smooth to $1$\% smooth (i.e. $0$\% to $99$\% clumpy).

\begin{figure}[h]
\includegraphics[scale=0.3]{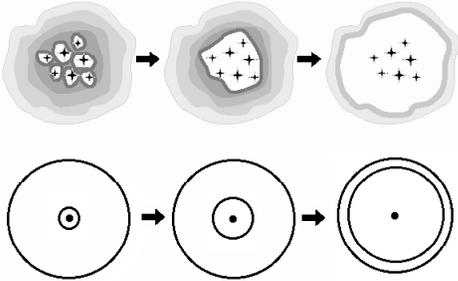}
\caption{\footnotesize{A cartoon illustrating the early stages of a super
star cluster's evolution and our corresponding model estimates.  The
smallest inner radius presented in this paper, $R_{\rm in}=0.1$~pc, is
likely to correspond to intra-cluster dust in a real cluster
(left-most figure), which will only exist while the constituent stars
still have individual cocoons.}
\label{ClusterEvolution}}
\end{figure}

The models mimic a cluster's evolution by moving the inner radius out
towards the outer radius in a geometric sequence;
Table~\ref{model_tree} lists the sequence for the three possible outer
radii we have chosen, and Figure~\ref{ClusterEvolution} shows how the
geometric progression roughly follows what is expected for a real
SSC. At very young ages, each star will realistically inhabit its own
compact H{\sc ii} region, but as the cluster evolves, these compact
regions will grow and merge into a single, large cavity. Assuming a
stationary stellar wind as in the semi-analytical and numerical models
presented by \citet{tenorio07}, this cavity will grow and disperse,
eventually leaving the central SSC exposed. The initial inner radius
of the models, correlated to intra-cluster dust, is $0.1$pc,
and moves out to the outer radius, which remains fixed.

The SFE values from $5$\% to $50$\% presume initial cocoon masses in
the range $1.9 \times 10^7$M$_\sun$ down to $1 \times
10^6$M$_\sun$. For a given SFE value and outer radius, the average
mass density is constant with radius, so that the mass of the shell in
a geometric sequence scales with the inner and outer radii as: $M
\propto R_{out}^3 - R_{in}^3$. The average mass densities range from
$10^5$ cm$^{-3}$ at the densest to $10$ cm$^{-3}$ at the least dense,
where the densest models are for a SFE of $5$\% and an outer radius of
$25$pc while the least dense models are for a SFE of $50$\% and an
outer radius of $100$pc.

The envelope is assumed to have a standard dust-to-gas ratio of
$0.01$. It is furthermore seeded with PAHs/VSGs to a depth equal to
$A_V = 1.5$ from the ionizing source. This is because the UV flux is
sufficiently attenuated after this depth that we expect very little UV
heating of small grains in the rest of the envelope. Comparisons
between models run with PAHs/VSGs throughout the dusty envelopes and
models PAHs/VSGs in to a depth of $A_V = 1.5$ show that PAH/VSG
emission changes only slightly for models where $A_V \lesssim 250$.

There are five clumpy dust fractions presented: $0$, $0.1$, $0.5$,
$0.9$, \& $0.99$ clumpy. The spread in clumpy fraction is the complete
range of conceivable clumpy values that could be found. The volume
fractal dimension was chosen as the best value from
\citet{elmegreen97} which matched the turbulent intercloud medium,
D=2.3. The fractal length L is taken to be 3.792 so that the maximum
density of the clumps can be computed per hierarchic level, $\rho(h) =
L^{(3-D)h+HD}$ where $h$ is the hierarchical level in question and $H$
is the maximum hierarchical level, which we have set equal to five;
higher $H$ values create fractal sizes smaller than the resolution
elements in our three-dimensional grid.

\subsection{Supporting Observational Evidence for the Model Parameter Choices \label{ObsEvidence}}

The model parameters were chosen to be consistent with observations of
embedded SSCs, and the parameter space is corroborated with
observational evidence from the literature in this section.

With regard to dust grains types, the PAH models from
\citet{draineli07} are the most suitable available to date. Ongoing
research \citep{draineli07, galliano08} that accounts for different
PAH ionization fractions and redder exciting sources than those used
here would likely change individual PAH band fluxes. However, since
we are not interested in PAH ratios in this work, the inclusion of
these processes does not impact the results of this study.

The ice mantle on the dust grains is suitable for dense, cold regions,
such as young star formation regions, where a large part of the dust
is not being heated by the embedded stars and the water-ice cannot be
sublimated from the grains. At any optical depth it has been shown
that the water-ice features will be masked and therefore hard to
detect \citep{robinson10}. Since there is observational evidence for
water associated with compact, young star clusters \citep{brogan10}
and the slope of the opacity functions with and without water ice are
the same, we have left the dust with ice mantles throughout the entire
geometric sequence.

Fixing the cluster age is based on upper limits of the time it takes
for a cluster to emerge from its envelope. Observations with HST show
optically visible clusters with ages of just a few Myr
\citep[e.g.][]{johnson00}. Radio observations have also suggested the
same approximate age, so we can assume a fiducial emergence timescale
of about three Myr for a super star cluster \citep{whitmore02,
kobulnicky99}. Since the UV continuum, which most affects the
resulting infrared SED, does not significantly change between ages of
$0$ and $~3$-$4$ Myr, setting the cluster age to $1$ Myr throughout
the sequence is warranted in order to limit the number of free
parameters.

The models are run under the assumption that the central star cluster
is a point source and the envelope is perfectly spherical; this is a
necessary simplification to confine the parameter space in this
study. However, the high stellar densities and small half-light radii
discovered for optically-visible SSCs \citep{tacconi96, demarchi97}
suggest that the central source is fairly localized in its formation
envelope. Furthermore, placing dust to within $0.1$pc of the central
point source will produce the high dust temperatures expected from
intracluster dust in a real embedded SSC. Dust sublimation is not a
concern even for these most embedded model phases: the dust
sublimation radius, assuming a sublimation temperature of $1600$K, is
$0.03$pc, well within the smallest inner radius value adopted.

The outer radii of $25$, $50$, and $100$pc were chosen based on
estimates from observational results. \citet{vacca02} model infrared
and radio observations of `ultra-dense' H{\sc ii} regions in Henize
2-10 as scaled-up versions of Galactic ultra-compact H{\sc ii} regions
to derive their radial size. \citet{hunt05} based their estimates on
models of global dwarf galaxy SEDs for which the radii of the actively
star-forming regions are derived based on infrared luminosities and
temperatures.

Lastly, in order to be consistent with the growing body of evidence
that suggests that the ISM is clumpy in a large variety of
environments, a clumpy envelope was included around the central star
cluster. That the ISM is not uniform is well established
\citep[e.g.][]{verschuur95,elmegreen97b,kim08}. In star formation
environments, recent work has begun to show that the same is
true. \citet{thuan05} shows that the molecular hydrogen in a BCD
galaxy is clumpy because emission is visible in the near-IR, which is
sensitive to dense H$_2$, but not in the UV, which is sensitive to
diffuse H$_2$. \citet{reines08} and \citet{johnson09} show that for
different clusters, more than $40$\% and $50$\% of the UV flux is detected
outside the embedded sources. This could either be because the ISRF is
very strong around the clusters or because there is light leaking from
the clusters. Most recently, model-dependent results by
\citet{eldridge10} shows that about $50$\% of the ionizing flux is
leaving the giant H{\sc ii} region NGC$604$, though this latter result
is highly dependent on ages derived from observations of Wolf-Rayet
stars.

\section{RESULTS \label{results}}

\subsection{Simulated Images \label{images}}

\begin{figure}[h]
\includegraphics[scale=0.4]{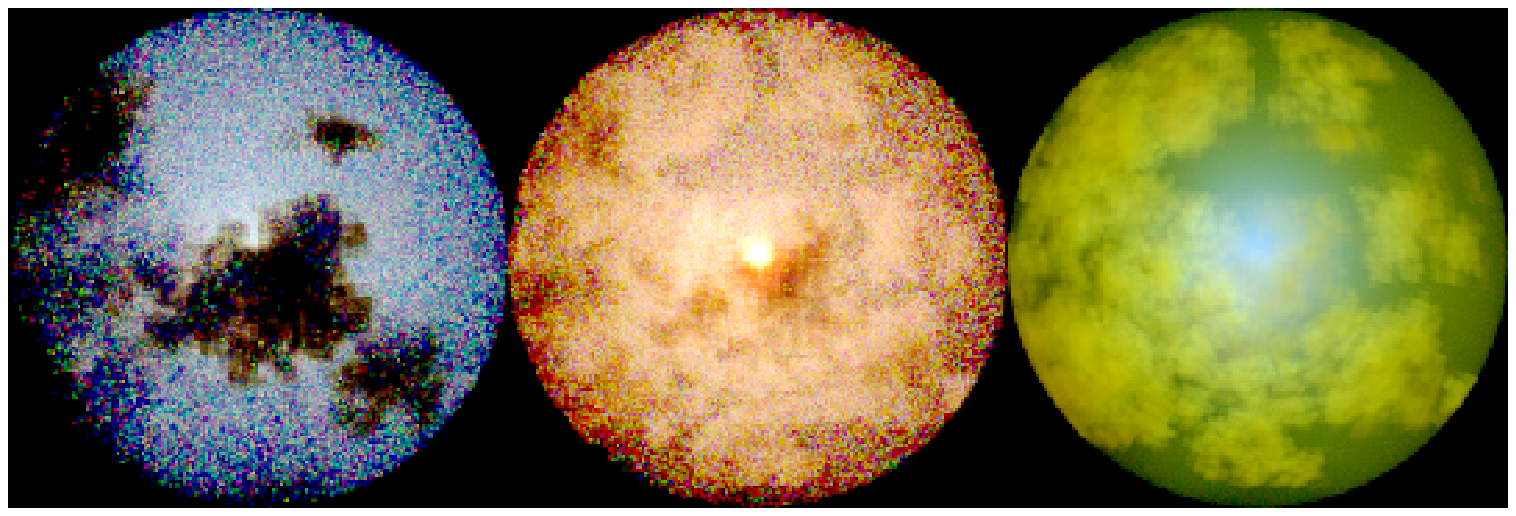}
\includegraphics[scale=0.4]{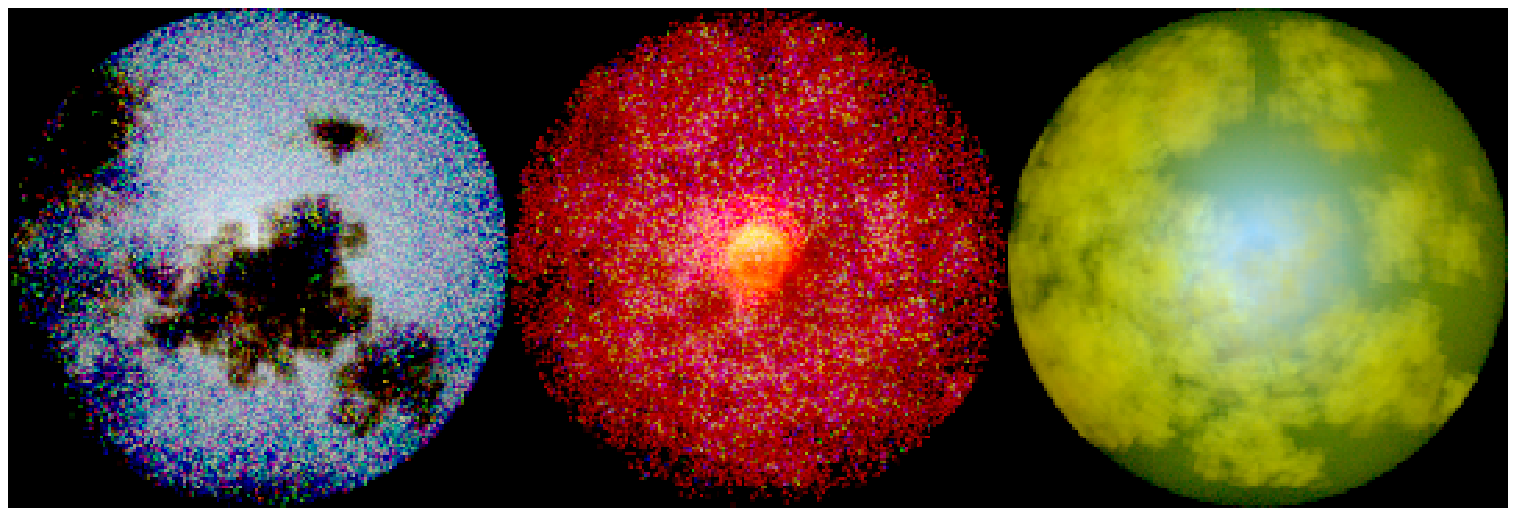}
\includegraphics[scale=0.4]{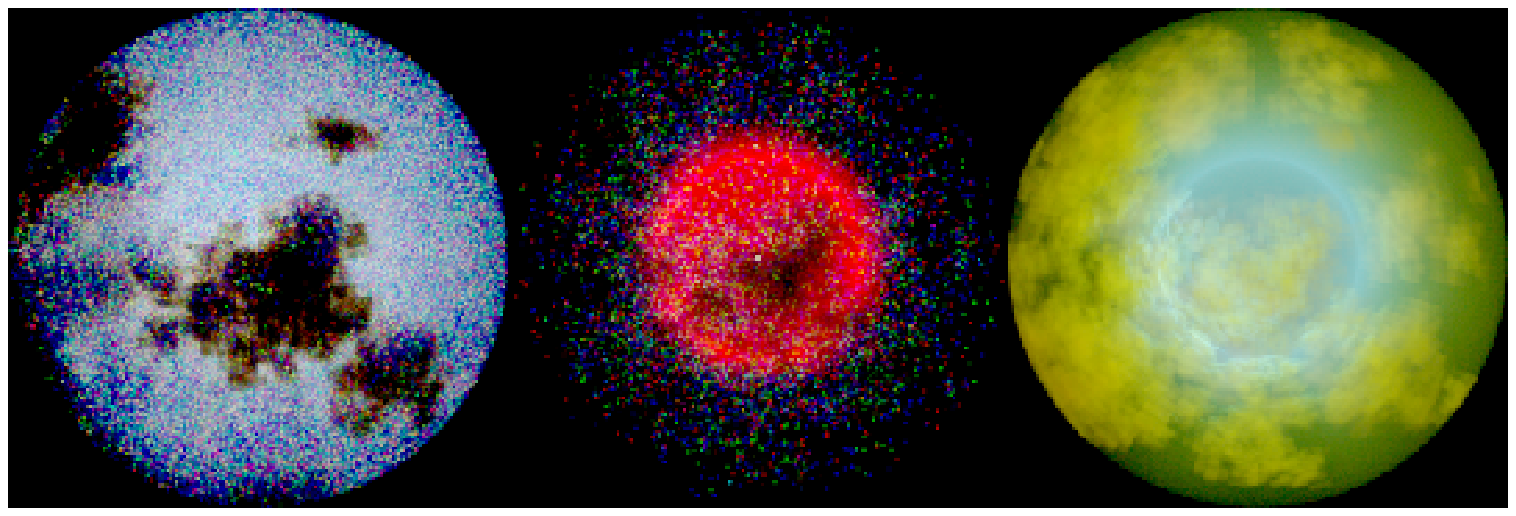}
\includegraphics[scale=0.4]{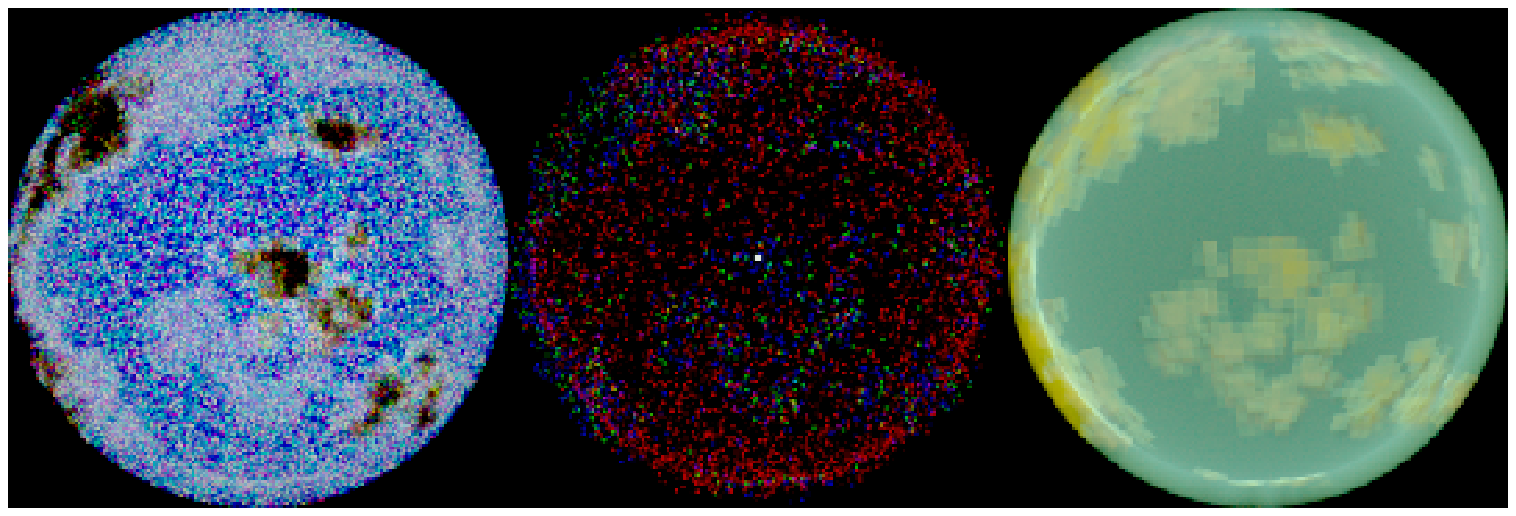}
\caption{\footnotesize{Simulated color images for a geometric
(pseudo-evolutionary) sequence of a natal SSC for a SFE of $10$\%,
outer radius~$=50$~pc, and clumpy fraction~$=90$\%.  Images are shown
for an inner radius of the cocoon of $1$~pc, $5$~pc, $20$~pc, and
$45$~pc, from top to bottom respectively.  The colors shown are for
combinations of J, H,and K (left), IRAC $3.6$\micron, $4.5$\micron,
and $8$\micron~(middle), and MIPS $24$\micron, $70$\micron, and
$160$\micron~(right).}
\label{color_image}}
\end{figure}

Images of the clusters at several stages along the geometric sequence
are shown in Figure~\ref{color_image} in near-IR (J, H, and K bands),
mid-IR \citep[{\em Spitzer} IRAC 3.6\micron, 4.5\micron, and
8.0\micron~ bands;][]{fazio04}, and far-IR \citep[{\em Spitzer} MIPS
24\micron, 70\micron, and 160\micron~bands;][]{rieke04} light from
left to right. They were created from the SFE~$=10$\% models with an
outer radius of $50$pc, clumpy dust fraction of $0.9$ and inner radii
of $1$, $5$, $20$, and $45$pc from top to bottom.

One of the striking features of these simulations is that a
significant amount of the near-IR light can escape the clumpy
envelope, as seen by the scattered light on the inside of the cloud
surfaces. Due to the scattered light, the near-IR images appear blue
in color, because scattering favors bluer photons (in this case,
J-band photons are scattered more regularly than K-band photons). The
mid-IR band images, sensitive to warm dust and PAH/VSG emission,
appear red because of the dominant PAH feature and high dust continuum
in the 8\micron~band. As the inner radius moves outward, the MIR
colors become more red because the inner dust temperature becomes
lower, removing what little hot dust radiation in the small-wavelength
bands there was. Additionally the hole appears to be the brightest
feature in the mid-IR because the shell is being lit up by the central
source. The far-IR bands probe both sides of the infrared dust peak,
with the 24\micron~ emission showing hotter dust and 70 and
160\micron~ bands showing cold dust emission. Since the dust peak is
typically near 70\micron, these images therefore appear green.

\subsection{The Change in Ultraviolet Flux with Clumpy Dust Fraction \label{UV}}

\begin{figure}[h]
\includegraphics[scale=0.4]{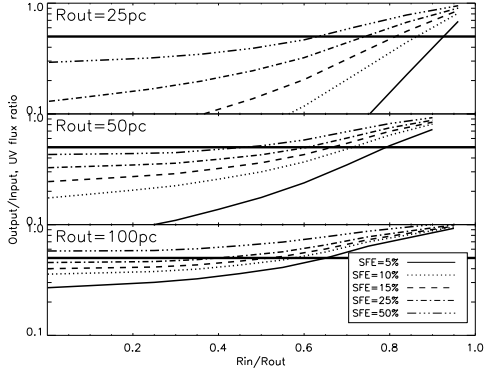}
\caption{\footnotesize{The fraction of UV light that escapes from the $99$\%
clumpy embedding envelope as a function of R$_{in}$/R$_{out}$. The
horizontal line is at a value of $0.5$, signifying $50$\% of the UV
flux is not being reprocessed by the dust, and is only meant to help
guide the eye. The solid line is for a SFE of $5$\%, dotted line for
$10$\%, dashed line for $15$\%, dash-dot line for $25$\%, and
dash-dot-dot for $50$\%.}
 \label{UVLight}}
\end{figure}

Motivated by observations suggesting that roughly $50$\% of the UV
flux leaks from embedded star clusters \citep[see][for details on the
observations]{reines08,johnson09}, we plot the fraction of UV light
lost as a function of the R$_{in}$/R$_{out}$ ratio for the $99$\%
clumpy models in Figure~\ref{UVLight}.

Figure~\ref{UVLight} shows that R$_{in}$/R$_{out}$ ratios $\gtrsim
0.5$ generally reproduce the results, depending on the SFE and
R$_{out}$ values adopted. The very high SFE values, such as $50$\%,
always allows a large fraction of the ultraviolet photons out of the
envelope because of its low optical depth. The clumpiness will also
have an effect on the fraction of UV light leakage, and generally less
light will leak from envelopes with smaller clumpy fractions. 

Since the radio sources studied in \citet{reines08} and
\citet{johnson09} are obscure, however, their findings are probably
for low-SFE, moderately evolved regions where the optical extinction
for a corresponding model is about A$_V \thicksim 30-50$ on
average. The UV light leakage would therefore take place along those
sightlines exhibiting minimal extinction, A$_V \lesssim 1$.

\subsection{Infrared Variation with Viewing Angle \label{ViewingAngle}}

The clumpy dust distribution creates variations in infrared luminosity
and certain spectral features depending on the viewing angle. In a
general sense, clumpier media show the most variation in spectral
features and luminosity compared to smoother media.

\begin{figure}[h]
\includegraphics[scale=0.3,angle=90]{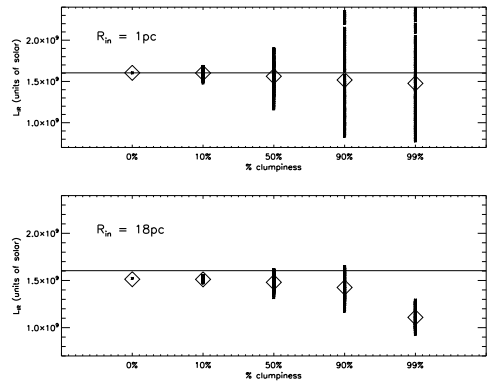}
\caption{\footnotesize{The computed infrared luminosity (3-1000\micron)
plotted versus the fraction of dust that is in clumps for an optically
thick (top) and an optically thin (bottom) model. R$_{out}$ is $25$pc
for both panels. The horizontal line is the input stellar cluster
luminosity, which matches the $L_{IR}$ for optically thick models. The
$L_{IR}$ computed along each sightline, assuming a spherical, smooth
distribution, is plotted with stars, while the $L_{IR}$ values
determined by adding up all sightlines for a model are shown as
diamonds (this is the correct infrared luminosity). The clumpier
models, as well as showing a larger dispersion in computed infrared
luminosity values, also exhibit lower infrared luminosities in
general, because the clumpy envelope allows ultraviolet photons to
exit the envelope (see Figure~\ref{UVLight}).}
\label{LirVar}}
\end{figure}

Figure~\ref{LirVar} demonstrates how the infrared luminosity derived
from a random sightline will be incorrect due to a clumpy
envelope. The infrared luminosity inferred from a single sightline
observation, plotted as stars in the figure and defined as the
$3$-$1000$\micron~ integrated luminosity, can vary from roughly half
to nearly twice the true infrared luminosity for a clumpy dust
distribution. The true infrared luminosities are plotted as diamonds
and are the infrared luminosities summed over all viewing angles for a
model. For smooth dust distributions with thick envelopes, the input
central source luminosity is equal to the true infrared
luminosity. The infrared luminosities are compared to the stellar
cluster's input luminosity of $1.604 \times 10^9$L$_\sun$, plotted as
the solid horizontal line.

\begin{figure}[h]
\includegraphics[scale=0.3]{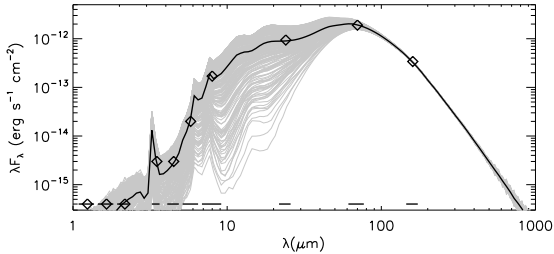}
\caption{\footnotesize{The variation of the SED for a given model due to the
viewing angle of the observer.  For this example, the model shown has
R$_{\rm out}=25$~pc, R$_{\rm in}=3$~pc, a clumpy fraction of 0.90, and
a 10\% SFE.  The mean SED is shown with a black line, and the near-IR
and {\it Spitzer} observing bands are indicated with diamonds, with
$\Delta \lambda$ for each band given at the bottom.}
\label{variation}}
\end{figure}

If a clump is along the line of sight in front of the central source,
the optical depth will appear to be high and the $9.8$ and
$17$\micron~ silicate absorption features will be deeper in the
infrared spectrum. However, clumps that are behind the central source
will reflect light, and silicate emission features will therefore be
evident. This is illustrated in Figure~\ref{variation}, which shows
that the features of the infrared spectrum of a source inside a clumpy
envelope are highly sightline-dependent. The average spectrum is also
shown for comparison.

\subsection{Spectral Energy Distribution Properties \label{SEDs}}

\begin{figure}[h]
\includegraphics[scale=0.3]{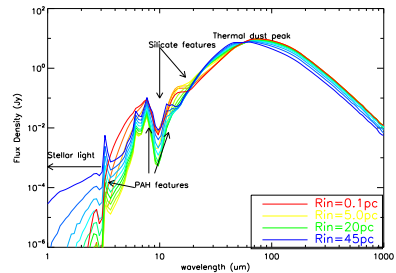}
\caption{\footnotesize{The spectra shown are along the geometric sequence in
Table~\ref{model_tree} for R$_{out}=50$pc and a smooth dust
distribution. As R$_{in}$ increases towards R$_{out}$, the silicate
features, most noticeably the one at $\sim 9.8$\micron, become
shallower. The PAH features are about the same throughout the
sequence. The envelopes with small inner radii are more opaque to
short-wave radiation and reprocess it to longer wavelengths.
Therefore less flux between $1$-$5$ \micron~ is measured and more flux
at $100$ \micron, shifting the peak of the SED to the red. The stellar
light is less extincted at larger R$_{in}$ values.}
\label{sfe05pc50smooth}}
\end{figure}

\begin{figure}[h]
\includegraphics[scale=0.35]{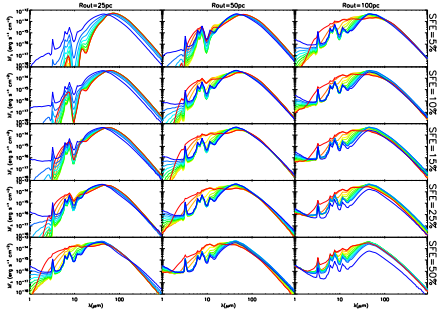} 
\caption{\footnotesize{The infrared SEDs of all of the geometric sequences
using a smooth dust distribution are shown; the range in R$_{in}$
values from small to large are as in Figure~\ref{sfe05pc50smooth}.}
\label{SmoothSpectralMatrix}}
\end{figure}

\begin{figure}[h]
\includegraphics[scale=0.35]{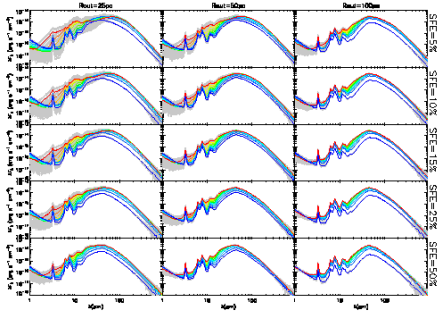} 
\caption{\footnotesize{The infrared SEDs of all the geometric sequences using
a $99$\% clumpy dust distribution, with the sightline-average data
shown in colors as in Figures~\ref{sfe05pc50smooth}~and
\ref{SmoothSpectralMatrix} and the individual sightline data shown in
gray scale.}
\label{ClumpySpectralMatrix}}
\end{figure}

The output spectral energy distributions all generally have four
components: ($1$) an extincted stellar spectrum from the central
source; ($2$) thermal dust emission that peaks between $40$ and
$70$\micron; ($3$) silicate emission or absorption features; and ($4$)
the PAH features visible in the mid-IR. Figure~\ref{sfe05pc50smooth}
shows the smooth dust SEDs for a geometric sequence as an example of
how these four main components are expected to change as the cluster
evolves. The geometric sequences for all SFE and R$_{out}$ values are
given in Figures~\ref{SmoothSpectralMatrix} and
\ref{ClumpySpectralMatrix} for the smooth and $99$\% clumpy dust
distributions. In Figure~\ref{ClumpySpectralMatrix}, the gray spectra
are individual sightline SEDs for each R$_{in}$ value, while the
colors are the sightline-averaged values\footnote{The suite of model
SEDs and photometric bands along all viewing angles are available via
D.~G. Whelan's webpage:
$http://www.astro.virginia.edu/\sim dww7v/SD\_models/$}.

For the most dense models, including low SFE values and small R$_{in}$
values, the starlight is largely absorbed by the envelope. For clumpy
envelopes, many sightlines allow optical and UV photons to pass
through so that a stellar continuum is visible, while the clumps have
very high extinction values. For smooth dust distributions the
starlight is largely absorbed along all sightlines. The A$_V$ values
for smooth distributions compared to the average values and ranges
found in clumpy solutions, are listed in Table~\ref{AvValues} for the
R$_{in}=0.1$pc models. The average A$_V$ values for evolved stages in
the geometric sequence are relatively small for all models (roughly
less than $2$), hence the spectra of clumpy and smooth models at high
SFE value and large inner radii are similar.

The wavelength of the far-IR peak of the thermal grain emission
depends on the predominant temperature of the dust in a model
envelope. For the high column density envelopes, which have low SFE
values and small R$_{in}$ values, the UV light is absorbed in the
inner portions of the envelope so that much of the envelope remains
quite cold. Therefore, low SFE values and small R$_{in}$ values tend
to produce infrared peaks at about $60$\micron. For the optically thin
models with high SFE values and large inner radii, a large proportion
of the dust is being heated by the cluster, so the far-IR peak moves
to shorter wavelengths.

Silicate absorption is directly proportional to the optical depth, so
that the deepest silicate features appear in the largest column-depth
models. Silicates in emission, as was described in
\S~\ref{ViewingAngle}, appear in clumpy models where sightlines
include dust clumps illuminated behind the central source, which
re-emit absorbed starlight as $10$\micron~ light that is unblocked by
any intervening clouds.

The PAH features are almost ubiquitous in the model SEDs. The early
stages in the geometric sequence appear to have reduced emission from
the PAHs for two reasons. The first reason is that for the high
density models, there is very little volume from which the PAHs can be
excited; PAHs are only excited by UV photons to a depth of
A$_V=1.5$. The second reason is because the thermal hot dust component
dominates the mid-IR flux for the early stage models, which lowers the
measured equivalent widths of these features. For later stages on the
geometric sequence, the PAH emission is relatively constant along the
sequence because as the ultraviolet flux on the inner surface of the
envelope decreases as r$^{-2}$, the surface area available to excite
PAH molecules increases as r$^2$.

\subsection{Infrared Colors as Tracers of the Input Parameters \label{colors}}

The model SEDs have been convolved with numerous filter sets for
comparison to photometric observations. These include the four IRAS
bands ($12$, $25$, $60$ and $100$\micron), three standard near-IR
bands (J, H, and K), {\em Spitzer} IRAC bands ($3.6$, $4.5$, $5.8$,
and $8.0$\micron), {\em Spitzer} MIPS bands ($24$, $70$, and
$160$\micron), {\em Herschel} PACS bands($70$, $100$, and
$160$\micron) and {\em Herschel} SPIRE bands ($250$, $350$, and
$500$\micron).

\begin{figure}[h]
\includegraphics[scale=0.4]{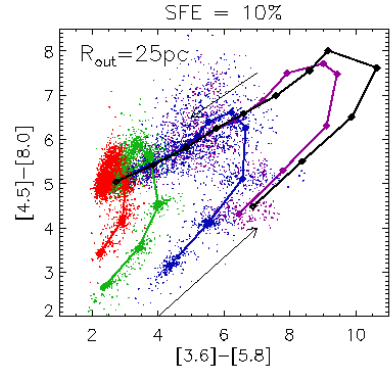}
\caption{\footnotesize{Using the mid-IR Spitzer/IRAC colors, the geometric
sequence for R$_{out}=25$pc and SFE$=10$\% is shown for smooth models
(black line), $10$\% clumpy models (purple), $50$\% clumpy models
(blue), $90$\% clumpy models (green), and $99$\% clumpy models
(red). The geometric sequences begin towards the bottom with
R$_{in}=0.1$pc and follow the arrows around the turn-around to their
termination at R$_{in}=24$pc. The solid lines and diamonds trace out
the sightline-averaged colors, and the dots represent individual
sightlines. While even the individual average tracks often overlap,
the different sightlines add a significant uncertainty to any
interpretation of these data.}
\label{DegenerateColors}}
\end{figure}

As an example of how colors can be used to plot the data, we show the
IRAC colors for the R$_{out}=25$pc and SFE$=10$\% models for all
clumpiness values and all R$_{in}$ values in
Figure~\ref{DegenerateColors}. There are degeneracies in the model
average values at late evolutionary stages, and the individual
sightlines make the interpretation of data extremely difficult.

\begin{figure}[hp]
\includegraphics[scale=0.35]{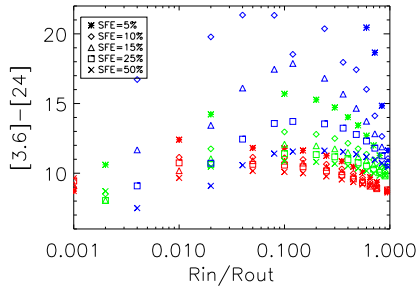}
\caption{\footnotesize{The [$3.6$]-[$24$] color is plotted versus the ratio
of the inner and outer envelope radii for the three different outer
radius values considered, $25$, $50$, and $100$pc (shown in blue,
green, and red, respectively) using the smooth dust distribution
models only. If the star formation efficiency (SFE) can be determined
using the [$70$]-[$160$] color as shown in
Figure~\ref{BreakDegeneraciesSmooth2}, then the R$_{in}$/R$_{out}$
value can be estimated using this plot.}
\label{BreakDegeneraciesSmooth1}}
\end{figure}

\begin{figure}[hp]
\includegraphics[scale=0.35]{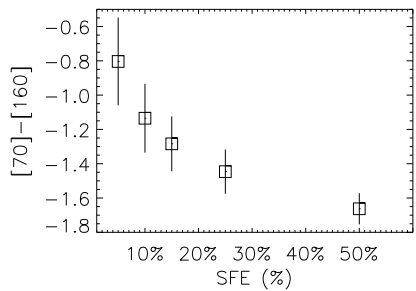}
\caption{\footnotesize{The [$70$]-[$160$] color is used to separate different
SFE values for the models with a smooth dust distribution. The bars
shown are the dispersion and there is a fair bit of overlap between
the models. One can only expect, therefore, to be able to roughly
determine the SFE for the smooth models.}
\label{BreakDegeneraciesSmooth2}}
\end{figure}

Because the clumpy dust distributions create degeneracies in the SEDs,
diagnostics to recover the input parameters were developed using the
smooth dust distribution models first. In
Figure~\ref{BreakDegeneraciesSmooth1} the [$3.6$]-[$24$] color is
plotted against the R$_{in}$/R$_{out}$. Red, green, and blue are used
to designate models with different R$_{out}$ values (red, green, and
blue for $100$, $50$, and $25$pc, respectively), and the plotting
symbols are for the five different SFE values considered. Although
there is significant overlap between the colors, the tracks are
separate over most of the R$_{in}$/R$_{out}$ values for each
individual SFE value. The [$70$]-[$160$] color can be used to roughly
determine the SFE value as shown in
Figure~\ref{BreakDegeneraciesSmooth2}; the bars represent the model
dispersions. There is significant overlap between dispersion in SFE
values of $10$\% to $25$\%, but [$70$]-[$160$] values of $> 0.9$
represent very low SFE values ($< 10$\%), values between $\sim -1.0$
and $-1.5$ represent SFE values between $10$\% and $25$\%, and values
$> -1.6$ will be for SFE of $50$\% or greater.

Colors that can best be used to recover the input parameters for the
entire, clumpy suite of models were also investigated. Given that the
value of the input parameters and the viewing angle can both
significantly change an embedded source's SED and colors, there is a
need to determine which, if any, photometric measurements can be used
to reliably constrain the physical geometry. For a given input
parameter (SFE, clumpiness, R$_{in}$, or R$_{out}$), the following
calculation was made: for each color, the mean color and standard
deviation was measured with the input parameter fixed and everything
else (i.e. the other input parameters and all viewing angles)
variable. This was done at each fixed input parameter value (e.g. for
each of the five different SFE values). Finally the difference in the
means is divided by the greatest of the standard deviations - this is
a measure of how much the input parameter affects the color, compared
to the other input parameters {\em and the viewing angle ambiguities}.

This analysis is related to a Principal Component Analysis, but
adjusted to our goal of finding the minimal set of colors with maximal
{\em physical} diagnostic power. Rather than solving freely for
eigenvectors of the model set in color space, which may not directly
correspond to the physical variables, we hypothesize that a principal
component exists for which the eigenvalue would directly correspond to
a physical parameter, and then measure the projection of that
hypothetical component onto each of our color axes. This process
allows of course that there is no such `physically diagnostic
principal component,' in which case the variation of all colors with
the physical parameter would be small compared to the standard
deviation in the color due to other causes.

\begin{figure}[hp]
\includegraphics[scale=0.35]{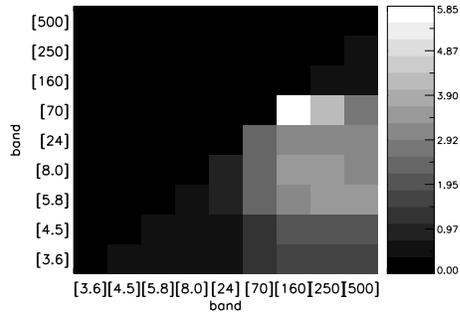}
\caption{\footnotesize{Determining the best color (y-axis magnitude minus
x-axis magnitude) for recovering the star formation efficiency
(SFE). In order to determine the best color for discriminating between
different SFE values, this calculation was made: the spread in the
mean color values between the different SFE values was divided by the
greatest standard deviation of each SFE value was computed. The
maximum resulting value is the most discriminating color (see
\S\ref{colors}). Each color is the y-axis magnitude minus x-axis
magnitude, and the best color was found to be [$70$]-[$160$]. How well
this `best' color helps to determine the SFE is shown in
Figure~\ref{BreakDegeneraciesAll1}, and discussed fully in
\S\ref{colors}.}
\label{ChooseColorsSFE}}
\end{figure}

\begin{figure}[hp]
\includegraphics[scale=0.35]{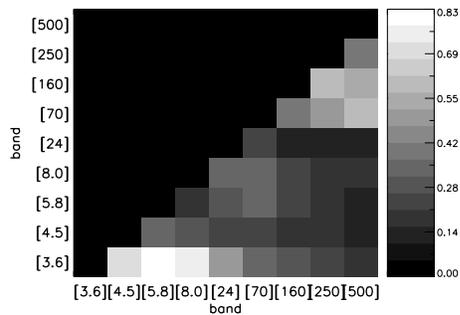}
\caption{\footnotesize{Determining the best color for recovering the fraction
of dust in a smooth distribution in the envelope. Using the criterion
described in \S\ref{colors}, the best color for recovering the
percentage of dust that is smooth was found to be
[$3.6$]-[$5.8$]. However, as shown in
Figure~\ref{BreakDegeneraciesAll2}, even this `best' color cannot be
used to discriminate between the average values.}
\label{ChooseColorsSmthFract}}
\end{figure}

\begin{figure}[hp]
\includegraphics[scale=0.35]{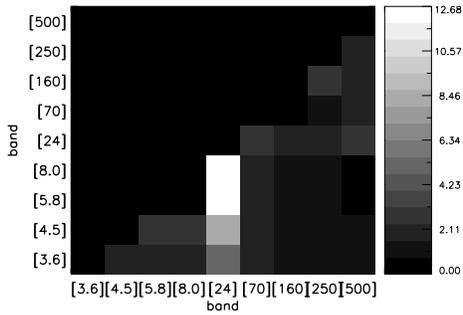}
\caption{\footnotesize{Determining the best color for recovering the
R$_{out}$ ratio. The best discriminator found for R$_{out}$, which
also well separates and R$_{in}$/R$_{out}$ ratios, is [$8.0$]-[$24$],
and is shown in Figure~\ref{BreakDegeneraciesAll3}.}
\label{ChooseColorsRiRo}}
\end{figure}

Figures~\ref{ChooseColorsSFE} to \ref{ChooseColorsRiRo} are plots of
infrared color versus infrared color where the grayscale shows the
value of the quotient for determining the best diagnostic color
described above. Values in the figures near zero indicate little
spread in the average color for each input parameter value, while
values at the maximum have relatively large spreads in average color
values across the input parameter values, as well as small standard
deviations from the mean. These are therefore the best colors for
separating input parameter values: ($1$) [$70$]-[$160$] for SFE
(Figure~\ref{ChooseColorsSFE}); ($2$) [$3.6$]-[$5.8$] for fraction of
dust in a clumpy distribution (Figure~\ref{ChooseColorsSmthFract});
and ($3$) [$8.0$]-[$24$] for both R$_{out}$ and the R$_{in}$/R$_{out}$
ratio (Figure~\ref{ChooseColorsRiRo}).

\begin{figure}[hp]
\includegraphics[scale=0.35]{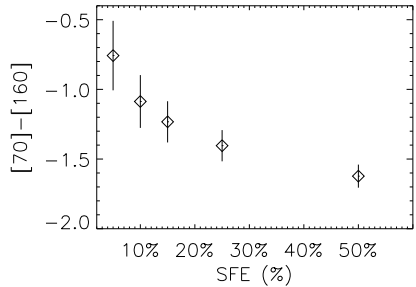}
\caption{\footnotesize{The [$70$]-[$160$] color that can best be used to
recover the SFE value. The diamonds are the average values over the
entire dataset, and the error bars represent the one-$\sigma$ errors
at each SFE value.}
\label{BreakDegeneraciesAll1}}
\end{figure}

\begin{figure}[hp]
\includegraphics[scale=0.35]{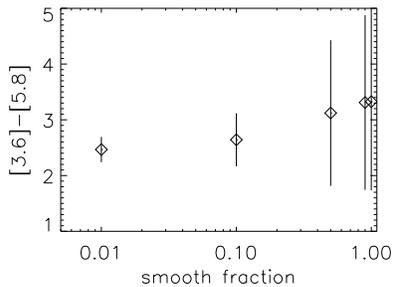}
\caption{\footnotesize{The [$3.6$]-[$5.8$] color that was found to best
recover the fraction of dust that is smoothly distributed in the
envelope. Given the 1-$\sigma$ error bars for the smoothest models
(i.e. there is such a great diversity in this color due to the input
parameters) it is only possible to separate the smooth from the clumpy
models in a statistical sample where the standard deviation in the
color could be measured.}
\label{BreakDegeneraciesAll2}}
\end{figure}

\begin{figure}[hp]
\includegraphics[scale=0.35]{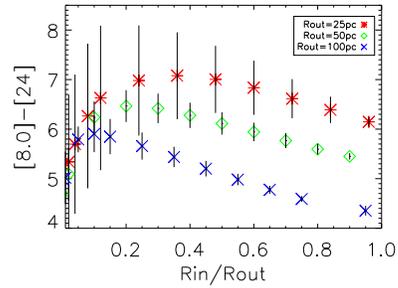}
\caption{\footnotesize{For envelopes with R$_{in}$/R$_{out} \gtrsim 0.4$ the
[$8.0$]-[$24$] color can be used to determine both the R$_{out}$
value, because they are well separated from each other, and the
R$_{in}$/R$_{out}$ ratio.}
\label{BreakDegeneraciesAll3}}
\end{figure}

Figures~\ref{BreakDegeneraciesAll1}, \ref{BreakDegeneraciesAll2}, and
\ref{BreakDegeneraciesAll3} plot these most useful colors' average
values and standard deviations for the whole suite of models versus
the input parameters. For SFE (Figure~\ref{BreakDegeneraciesAll1}), as
with the smooth dust distribution models, there is overlap in the
[$70$]-[$160$] color between the various SFE values. The overlap is
still most significant between $10$ and $25$\%, for which color values
between $-1.5$ and $-0.9$ would all be valid. Color values above
$-0.9$ match the $5$\% SFE, and color values below $-1.5$ match the
$50$\% SFE. For the smooth dust fraction, within the error bars in the
[$3.6$]-[$5.8$] versus smooth dust fraction plot
(Figure~\ref{BreakDegeneraciesAll2}) the color values are still
degenerate, and using thermal radio data \citep[such as was presented
in ][]{reines08} is a more reliable metric for determining the
fraction of ultraviolet light lost from an embedded star
cluster. However, because the standard deviations in color of the
clumpiest fractions are small compared to the smooth models (due to
the large differences in the input parameter values being explored), a
large, statistical study of embedded star clusters, for which average
color values and standard deviations could be computed, could
potentially be used to discern between the different smooth dust
fraction values. For R$_{in}$/R$_{out}$
(Figure~\ref{BreakDegeneraciesAll3}), colors are still intractable for
small R$_{in}$/R$_{out}$ values. For R$_{in}$/R$_{out} > 0.4$, the
[$8.0$]-[$24$] can be used both to distinguish between different
R$_{in}$/R$_{out}$ values and to distinguish between the different
R$_{out}$ values.

\clearpage

\section{COMPARISONS TO OBSERVATIONS \label{comparison}}

In the Milky Way, individual embedded massive stars are surrounded by
what have been identified as ``Ultracompact H{\sc ii} Regions''
(UCH{\sc ii}s), and it is hypothesized that the constituent massive
stars in embedded SSCs will also be surrounded by UCH{\sc ii}
regions. For this reason we compare our models to a sample of resolved
embedded massive stars in order to test whether our models have
reproduced, to first order, the necessary features of embedded massive
stars. The sample is an IRAS sample studied in \citet{wood89}, in
which it was found that 60\% of the brightest IRAS sources ($>
10^4$~Jy at 100\micron) in the color range of their galactic survey
are UCH{\sc ii} regions. \citet{wood89} also found that galactic
UCH{\sc ii} regions strictly obey a set of color criteria in the
infrared with $log(F_{60\micron}/F_{12\micron})>1.30$ and
$log(F_{25\micron}/F_{12\micron})>0.57$, while very few other types of
objects had IRAS colors fitting these criteria. Therefore, these color
criteria appear to be relatively robust for identifying UCH{\sc ii}
regions, and we might expect natal SSCs to obey these criteria as
well.

\begin{figure}[h]
\includegraphics[scale=0.45]{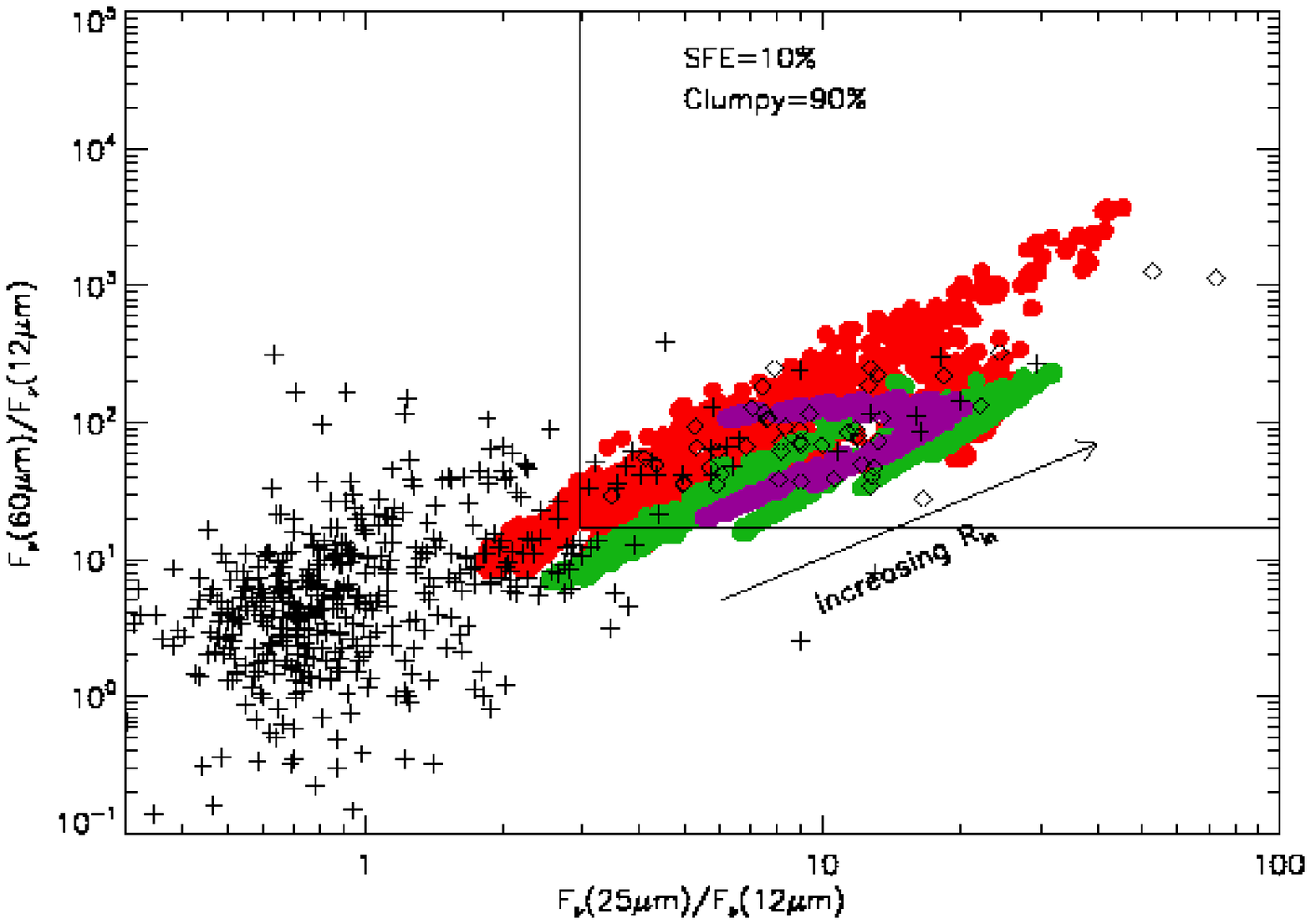}
\caption{\footnotesize{The IRAS colors from a subset of our models assuming
10\% SFE and clumpy fraction = 0.9 compared to field objects (plus
signs) and UCH{\sc ii} regions (diamonds) from the \citet{kurtz94}
survey.  Models with R$_{\rm out}=25$~pc are shown as red points,
R$_{\rm out}=50$~pc as green points, and R$_{\rm out}=100$~pc as
purple points.  The values of R$_{\rm in}$ used are those shown in
Table~\ref{model_tree}, and all sight lines are shown for each set of
model parameters.}
\label{plotWC}}
\end{figure}

We have convolved our model results with the IRAS filter profiles, and
Figure~\ref{plotWC} shows the IRAS colors from a subset of our models
compared to field objects (plus signs) and UCH{\sc ii} regions
(diamonds) from the \citet{kurtz94} survey (which includes the
\citet{wood89} sample).  In this figure, we focus on only the family of
models with a 10\% stellar mass component and a relative fraction of
clumpy dust of 90\%, although the other model families have IRAS
colors that are included in the range represented by the models
shown. Models with R$_{out}=25$~pc are shown as red points,
R$_{out}=50$~pc as green points, and R$_{out}=100$~pc as purple
points.  The values of R$_{in}$ used are those shown in
Table~\ref{model_tree}, and all sight-lines are shown for each set of
model parameters, which produces additional scatter.

There is good agreement between our models and the UCH{\sc ii}s sample
shown in Figure~\ref{plotWC}. There are a few of points to lower left
and upper right that are not traced by our models. The lower left
points have very hot dust in the inner portions of their envelopes,
and might be modeled with a smaller R$_{in}$ value that puts the hot
dust closer to the heating source. The points to upper right seem to
be an extension of the basic trend of increasing R$_{in}$. It is
therefore possible that we could reproduce these UCH{\sc ii}s if we
simply assumed a lower mass cutoff to our cluster which would lower
the UV flux incident on the inner envelope.

\section{KNOWN LIMITATIONS \label{limitations}}

Here we discuss some of the limitations of the models as a guide for
future improvements.

($1$) The models assume that the central star cluster is a point
source. While optical measurements have discovered half-light radii of
super star clusters of $1.7$pc on average \citep{demarchi97}, which is
significantly smaller than the assumed outer envelope radii, it is
nevertheless more realistic to place several ultracompact H{\sc ii}
regions at the center of the envelope (see
Figure~\ref{ClusterEvolution}). The intracluster dust would be heated
at early evolutionary stages if multiple, smaller clusters were used,
creating a hotter dust component than is seen with the current models.

($2$) Variations in metallicity have not been considered. For small
changes in metallicity (i.e. down to half or third solar metallicity),
we expect the dust-to-gas ratio to scale with metallicity. However,
more extreme environments, like blue compact dwarf galaxies, where
metallicities are often much lower than this, show dust in excess to
what is expected from their metallicity alone (T. X. Thuan, private
communication). Accounting for very low metallicities is therefore a
complicated issue that is probably best handled on an individual
basis.

($3$) The evolutionary sequences presented in this paper do not allow
the central star cluster to evolve. As described in
\S\ref{ObsEvidence}, the time between cluster formation and the
dispersal of the embedding envelope is short, on the order of $~3-4$
Myr. However, the dispersal rate is not well known, so we have not
made any attempt to place time stamps on the evolutionary sequence
presented here. Hydrodynamical models are needed to answer this
problem, and several hydrodynamical and semi-analytical studies of
super star clusters already exist \citep[e.g.][]{tenorio05, tenorio07,
wunsch07, wunsch08}. The rate at which the envelope is dispersed could
potentially be answered by similar studies of a super star cluster's
embedded phase.

($4$) This work has concentrated solely on the dust emission and
absorption in embedded super star clusters. Inclusion of nebular line
emission in the model spectra could be useful if line diagnostics
could be identified that break the model degeneracies.

($5$) The models do not include an interstellar radiation field (ISRF)
incident on the outside of the dusty envelope. An ISRF would heat the
dust on the outside of the envelope and provide an additional
component to the starlight visible in the UV/optical/near-IR
regime. This could affect the [$3.6$]-[$5.8$] color used to
distinguish between different fractions of clumpy dust.  The heating
of the outer envelope could affect the far-IR fluxes as well.

\section{SUMMARY \label{conclusions}}

We present SEDs and colors of embedded SSCs created using spherical
three-dimensional models. By varying the input parameters according to
a series of evolutionary sequences, we have created a suite of models
that can be used to constrain the evolutionary state of an embedded
super star cluster. The main conclusions of the study are:

($1$) A hierarchically clumped medium is suitable for recreating the
porous environments observed around embedded super star clusters
\citep[e.g.][]{johnson09};

($2$) The infrared luminosity derived from a single sightline
observation of a clumpy envelope can be wrong by as much as a factor
of two from the true value;

($3$) The infrared SED also varies with sightline in these clumpy
models, which has an impact on the near- and mid-infrared colors and
magnitudes, the strength of the observed silicate features, and the
dust continuum measured at these wavelengths;

($4$) For the smooth dust distribution models, the evolutionary
sequences that begin with mid-infrared obscure envelopes (A$_{V}
\gtrsim 50$) are marked by a gradual decline in the silicate absorption
features at $9.8$ and $17$\micron~ and a corresponding increase in the
visual and ultraviolet flux as the cluster envelope evolves. Those
sequences that begin with infrared-transparent envelopes (A$_{V} <
50$) instead have a predominant hot dust component and silicates in
emission at early stages that eventually both fade away as the inner
envelope radius moves outward. The clumpy envelope acts to confuse
these general trends, making it harder to determine envelope
properties.

($5$) Several diagnostic colors were found to constrain the envelope
properties. The {\em Spitzer} MIPS [$70$]-[$160$] color is found to be
a good diagnostic of the star formation efficiency, particularly at
separating very low and high values (such as $5$\% and $50$\%) from
more moderate values (between $10$\% and $25$\%). The [$3.6$]-[$5.8$]
color can be used to determine the fraction of clumpy dust in the
envelope for large samples of embedded super star clusters, but not
for individual sources (see \S\ref{colors} for details). In order to
determine the R$_{in}$ and R$_{out}$ values, the [$8.0$]-[$24$] color
can be used for R$_{in}$/R$_{out} \gtrsim 0.4$. Below this value the
data is degenerate for all colors.

($6$) The model IRAS colors trace the same area of color space as
ultracompact H{\sc ii} regions, the Galactic analogues to
extragalactic embedded super star clusters, suggesting that the models
will also be useful when data of resolved, embedded super star
clusters become available.

\acknowledgements DGW wishes to thank S. Ransom for the use of his
computer cluster to run the models. KEJ gratefully acknowledges
support for this paper provided by NSF through CAREER award 0548103
and the David And Lucile Packard Foundation through a Packard
Fellowship. All of the authors are grateful to the referee for their
help in improving the paper.

\clearpage

\begin{deluxetable}{ccc}
\tablecaption{The geometric sequences \label{model_tree}}
\tablehead{
\colhead{R$_{\rm out}=25$pc} & \colhead{R$_{\rm out}=50$pc} & \colhead{R$_{\rm out}=100$pc} \\ R$_{\rm in}$ & R$_{\rm in}$ & R$_{\rm in}$ }
\startdata
0.1pc & 0.1pc & 0.1pc \\
0.5pc & 0.5pc & 1.0pc \\
1.0pc & 1.0pc & 5.0pc \\
2.0pc & 5.0pc & 10pc  \\
3.0pc & 10pc  & 15pc  \\
6.0pc & 15pc  & 25pc  \\
9.0pc & 20pc  & 35pc  \\
12pc  & 25pc  & 45pc  \\
15pc  & 30pc  & 55pc  \\
18pc  & 35pc  & 65pc  \\
21pc  & 40pc  & 75pc  \\
24pc  & 45pc  & 95pc  \\
\enddata
\tablecomments{Each sequence is run using clumpy dust fractions of
  0.0, 0.1, 0.5, 0.9, and 0.99, and SFEs of 5\%, 10\%, 15\%, 25\%, and
  50\%, resulting in a total of 900 models. Stellar cluster mass of
  $10^6 M_\odot$, source luminosity $\sim1.6\times10^9 L_\odot$,
  Salpeter IMF $1-100M_\odot$, age 1 Myr, $Z_\odot$.}
\end{deluxetable}

\begin{deluxetable}{ccccccc}
\tabletypesize{\footnotesize}
\tablecaption{A table of average A$_V$ values and ranges for the
R$_{in}=0.1$pc models\label{AvValues}}
\tablehead{
\colhead{SFE(\%)} & \colhead{R$_{out}$} & \colhead{Smooth} & \colhead{$10$\%~clumpy} & \colhead{$50$\%~clumpy} & \colhead{$90$\%~clumpy} & \colhead{$99$\%~clumpy}}
\startdata
5  & 25  & 490  & 474  (441-569)   & 411  (245-885)   & 348  (49.1-1200) & 334  (4.92-1270)  \\
5  & 50  & 123  & 119  (111-143)   & 103  (61.4-221)  & 87.0 (12.3-300)  & 83.4 (1.24-318)   \\
5  & 100 & 30.8 & 29.8 (27.6-35.7) & 25.8 (15.4-55.4) & 21.8 (3.09-75.1) & 20.9 (0.310-79.5) \\
10 & 25  & 232  & 225  (209-270)   & 195  (116-419)   & 165  (23.2-569)  & 158  (2.34-603)   \\
10 & 50  & 58.2 & 56.3 (52.4-67.5) & 48.8 (29.1-105)  & 41.2 (5.83-142)  & 39.5 (0.587-151)  \\
10 & 100 & 14.6 & 14.1 (13.1-16.9) & 12.2 (7.30-26.2) & 10.3 (1.47-35.6) & 9.88 (0.147-37.7) \\
15 & 25  & 146  & 142  (132-170)   & 123  (73.2-264)  & 104  (14.6-358)  & 99.5 (1.48-379)   \\
15 & 50  & 36.7 & 35.5 (33.0-42.5) & 30.7 (18.3-66.1) & 26.0 (3.68-89.6) & 24.9 (0.369-94.8) \\
15 & 100 & 9.18 & 8.88 (8.27-10.7) & 7.69 (4.60-16.5) & 6.50 (0.925-22.4)& 6.23 (0.0925-23.7)\\
25 & 25  & 77.5 & 75.0 (69.7-89.9) & 64.9 (38.7-140)  & 54.9 (7.76-190)  & 52.7 (0.781-201)  \\
25 & 50  & 19.4 & 18.8 (17.5-22.5) & 16.3 (9.71-35.0) & 13.7 (1.95-47.4) & 13.2 (0.196-50.2) \\
25 & 100 & 4.87 & 4.71 (4.38-5.65) & 4.08 (2.44-8.76) & 3.44 (0.490-11.9)& 3.30 (0.0490-12.6)\\
50 & 25  & 25.8 & 25.0 (23.2-30.0) & 21.7 (12.9-46.6) & 18.3 (2.59-63.2) & 17.6 (0.260-67.0) \\
50 & 50  & 6.48 & 6.27 (5.83-7.52) & 5.43 (3.25-11.7) & 4.59 (0.652-15.8)& 4.40 (0.0652-16.7)\\
50 & 100 & 1.63 & 1.58 (1.47-1.89) & 1.36 (0.816-2.93)& 1.15 (0.163-3.96)& 1.10 (0.0163-4.20)
\enddata
\tablecomments{The R$_{in}$ value is $0.1$pc for all of these
values. The range in A$_{V}$ values depending on sightline are in
parentheses for the clumpy models, with the average values shown
first.}
\end{deluxetable}


\begin{thebibliography}{}

\bibitem[Ashman \& Zepf(2001)]{ashman01} Ashman, K.~M. \& Zepf, S.~E. 2001, 
\aj, 122, 1888

\bibitem[Beck et al.(2002)]{beck02} Beck, S.C., Turner, J.~L., 
Langland-Shula, L.~E., Meier, D.~S., Crosthwaite, L.~P., Gorjian, V. 2002, 
\aj, 124, 251

\bibitem[Bonnell et al.(2010)]{bonnell10} Bonnell, I. A., Smith,
R. J., Clark, P. C., \& Bate, M. R. 2010, \mnras, in
press. arXiv:1009.1152.

\bibitem[Bosch et al.(2009)]{bosch09} Bosch, G., Terlevich, E., \&
Terlevich, R. 2009, \aj, 137, 3437.

\bibitem[Brogan et al.(2010)]{brogan10} Brogan, C., Johnson, K., \&
Darling, J. 2010, \apj, 716, 51.

\bibitem[Cardelli et al.(1989)]{cardelli89} Cardelli, J.~A., Clayton,
G.~C., \& Mathis, J.~S. 1989, \apj, 345, 245

\bibitem[Clark et al.(2005)]{clark05} Clark, J. S., Negueruela, I.,
Crowther, P. A., \& Goodwin, S. P. 2005, \aap, 434, 949.

\bibitem[Chakrabarti \& Whitney(2009)]{chakrabarti09} Chakrabarti,
S. \& Whitney, B.~A. 2009, \apj, 690, 1432

\bibitem[Conti \& Vacca(1994)]{conti94} Conti, P. S., \& Vacca,
W. D. 1994, \apj, 423, L97

\bibitem[Cornette \& Shanks(1992)]{cornette92} Cornette, W.~M. \&
Shanks, J.~G. 1992 Applied Optics, 31, 3152

\bibitem[de Marchi et al.(1997)]{demarchi97} de Marchi, G., Clampin,
M., Greggio, L., Leitherer, C., Nota, A., \& Tosi, M. 1997, \apjl,
479, L27.

\bibitem[Draine et al.(2007)]{draine07} Draine, B.~T.,Dale, D.~A.,
Bendo, G., Gordon, K.~D., Smith, J.~D.~T., Armus, L., Engelbracht,
C.~W., Helou, G., Kennicutt, R.~C., Li, A., Roussel, H., Walter, F.,
Calzetti, D., Moustakas, J., Murphy, E.~J., Rieke, G.~H., Bot, C.,
Hollenbach, D.~J., Sheth, K. \& Teplitz, H.~I. 2007, \apj, 663, 866

\bibitem[Draine \& Li(2007)]{draineli07} Draine, B.~T., \& Li,
A. 2007, \apj, 657, 810

\bibitem[Eldridge \& Rela\~{n}o(2010)]{eldridge10} Eldridge, J. J., \&
Rela\~{n}o, M. 2010, \mnras, submitted. arXiv:1009.1871.

\bibitem[Elmegreen(1997)]{elmegreen97} Elmegreen, B. G. 1997, \apj, 477, 196

\bibitem[Elmegreen \& Efremov(1997)]{elmegreen97b} Elmegreen, B. G.,
\& Efremov, Y. N. 1997, \apj, 480, 235.

\bibitem[Fazio et el.(2004)]{fazio04} Fazio, G. G. et al. 2004, \apjs,
154, 10

\bibitem[Galliano et al.(2008)]{galliano08} Galliano, F., Madden,
S. C., Tielens, A. G. G. M., Peeters, E., \& Jones, A. P. 2008, \apj,
679, 310.

\bibitem[Holtzman et al.(1992)]{holtzman92} Holtzman, J. A., Faber,
S. M., Shaya, E. J., Lauer, T. R., Groth, E. J., Hunter, D. A., Baum,
W. A., Ewald, S. P., Hester, J. J., Light, R. M., Lynds, C. R.,
O'Neil, E. J., \& Westphal, J. A. 1992, \aj, 103, 691

\bibitem[Hunt et al.(2005)]{hunt05} Hunt, L., Bianchi, S., \& 
Maiolino, R. 2005, \aap, 434, 849

\bibitem[Indebetouw et al.(2006)]{indebetouw06} Indebetouw, R., 
Whitney, B. A., Johnson, K. E, \& Wood, K. 2006, \apj, 636, 362

\bibitem[Johnson et al.(1999)]{johnson99} Johnson, K. E., Vacca,
W. D., Leitherer, C., Conti, P. S., \& Lipsky, S. J. 1999, \aj, 117,
1708

\bibitem[Johnson et al.(2000)]{johnson00} Johnson, K.~E., Leitherer,
C., Vacca, W.~D., \& Conti, P.~S. 2000, \aj, 120, 1273

\bibitem[Johnson et al.(2003)]{johnson03a} Johnson, K. E., Indebetouw,
R., \& Pisano, D. J. 2003, \aj, 126, 101

\bibitem[Johnson \& Kobulnicky(2003)]{johnson03b} Johnson, K. E. \&
Kobulnicky, H. A.  2003, \apj, 597, 923

\bibitem[Johnson et al.(2009)]{johnson09} Johnson, K. E., Hunt, L. K.,
Reines, A. E. 2009, \aj, 137, 3788

\bibitem[Kim et al.(1994)]{kim94} Kim, S.-H., Martin, P. G., \&
Hendry, P. D. 1993, \apj, 422, 164

\bibitem[Kim et al.(2008)]{kim08} Kim, Y., Rieke, G. H., Krause, O.,
Misselt, K., Indebetouw, R., \& Johnson, K. E. 2008, \apj, 678, 287.

\bibitem[Kobulnicky \& Johnson(1999)]{kobulnicky99} Kobulnicky, H. A. \&
Johnson, K. E.  1999, \apj, 527, 154

\bibitem[Kurtz, Churchwell, \& Wood(1994)]{kurtz94} Kurtz, S.,
Churchwell, E., Wood, D.~O.~S. 1994, \apjs, 91, 659

\bibitem[Laor \& Draine(1993)]{laor93}
Laor, A. \& Draine, B. T. 1993, \apj, 402, 441

\bibitem[Leitherer et al.(1999)]{leitherer99} Leitherer, C., Schaerer,
D., Goldader, J. D., Gonz\'alez Delgado, R. M., Robert, C., Kune,
D. F., de Mello, D. F., Devost, D., Heckman, T. M. 1999, \apjs, 123, 3

\bibitem[Lucy(1999)]{lucy99} Lucy, L. B. 1999, \aap, 344, 282

\bibitem[Murray et al.(2010)]{murray10a} Murray, N., Quataert, E., \&
Thompson, T. A. 2010, \apj, 709, 191.

\bibitem[Murray et al.(2010)]{murray10b} Murray, N., Menard, B., \&
Thompson, T. A. 2010, \apj, submitted. arXiv:1005.4419

\bibitem[O'Connell et al.(1994)]{oconnell94} O'Connell, R. W.,
Gallagher, J. S., \& Hunter, D. A. 1994, 433, 65

\bibitem[Reines et al.(2008)]{reines08} Reines, A. E., Johnson, K. E.,
\& Hunt, L. K. 2008, \aj, 135, 2222

\bibitem[Rieke et al.(2004)]{rieke04} Rieke, G., et al. 2004, \apjs,
154, 25.

\bibitem[Robinson \& Maldoni(2010)]{robinson10} Robinson, G., \&
Maldoni, M. M. 2010, \mnras, 408, 1956.

\bibitem[Salpeter(1955)]{salpeter55} Salpeter, E. E. 1955, \apj, 121,
161

\bibitem[Sauvage \& Plante(2003)]{sauvage03} Sauvage, M., \& Plante,
S. 2003, \apss, 284, 941

\bibitem[Schweizer et al.(1996)]{schweizer96} Schweizer, F., Miller,
B. Y., Whitmore, B. C., \& Fall, S. M. 1996, \aj, 112, 1839

\bibitem[Shioya et al.(2001)]{shioya01} Shioya, Y., Taniguchi, Y., \&
Trentham, N. 2001, \mnras, 321, 11.

\bibitem[Sonneborn(2008)]{sonneborn08} Sonneborn, G. 2008, IAU
Symposium 250, 491

\bibitem[Tacconi-Garman et al.(1996)]{tacconi96} Tacconi-Garman,
L. E., Sternberg, A., \& Eckart, A. 1996, \aj, 112, 918.

\bibitem[Tenorio-Tagle et al.(2005)]{tenorio05} Tenorio-Tagle, G.,
Silich, S., Rodr\'{i}guez-Gonz\'{a}lez, A., Mu\~{n}oz-Tu\~{n}\'{o}n,
C. 2005, \apj, 620, 217.

\bibitem[Tenorio-Tagle et al.(2007)]{tenorio07} Tenorio-Tagle, G.,
W\"{u}nsch, R., Silich, S., Palou\u{s}, J. 2007, \apj, 658, 1196.

\bibitem[Thuan et al.(2005)]{thuan05} Thuan, T. X., Lecavelier des
Etangs, A., \& Izotov, Y. I. 2005, \apj, 621, 269.

\bibitem[Turner et al.(2000)]{turner00} Turner, J. L., Beck, S. C.,
\& Ho, P. T. P. 2000, \apj, 532, 109

\bibitem[Vacca, Johnson, \& Conti(2002)]{vacca02} Vacca, W. D.,
Johnson, K.~E., \& Conti, P.~S. 2002, \aj, 123, 772

\bibitem[Verschuur(1995)]{verschuur95} Verschuur, G. L. 1995, \apss,
227, 187

\bibitem[Warren(1984)]{warren84} Warren, S. G. 1984, \ao, 23, 1206

\bibitem[Whitmore(2002)]{whitmore02} Whitmore, B.~C. 2002, in: A Decade
of Hubble Space Telescope Science, M. Livio, K. Noll, M.
Stiavelli, Eds. (Cambride Univ. Press, Cambridge, 2002), pp. 153-180

\bibitem[Whitney et al.(2003)]{whitney03} Whitney, B. A., Wood, K.,
Bjorkman, J. E., \& Wolff, M. J. 2003, \apj, 591, 1049

\bibitem[Whittet et al.(2001)]{whittet01} Whittet, D. C. B., Gerakines,
P. A., Hough, J. H., \& Shenoy, S. S.  2001, \apj, 547, 872

\bibitem[Wood \& Churchwell(1989)]{wood89} Wood, D.~O.~S. \& Churchwell, E. 
1989, \apj, 340, 265

\bibitem[Wood et al.(2008)]{wood08} Wood, K., Whitney, B.~A., Robitaille, T., 
\& Draine, B.~T. 2008, \apj, 688, 1118

\bibitem[W\"{u}nsch et al.(2007)]{wunsch07} W\"{u}nsch, R., Silich,
S., Palou\u{s}, J., Tenorio-Tagle, G. 2007, \aap, 471, 579.

\bibitem[W\"{u}nsch et al.(2008)]{wunsch08} W\"{u}nsch, R.,
Teonorio-Tagle, G., Palou\u{s}, J., \& Silich, S. 2008, \aj, 135, 2222.

\end{thebibliography}
\end{document}